\documentstyle[eqsecnum,aps,prb]{revtex}
\begin{document}
\draft
\title{Spin pseudo--gap and interplane coupling in
Y$_{2}$Ba$_{4}$Cu$_{7}$O$_{15}$: \\
a $^{63}$Cu nuclear spin--spin relaxation study}

\author{R. Stern\cite{Auth1}, M. Mali, J. Roos and D.~Brinkmann}
\address{Physik-Institut, Universit\"at Z\"urich, CH--8057 Z\"urich,
Switzerland}
\date{October 28, 1994}
\maketitle
\begin{abstract}
We report measurements of the Gaussian contribution, $T_{2G}$, to the
plane $^{63}$Cu nuclear spin--spin relaxation time in the
YBa$_{2}$Cu$_{3}$O$_{7}$ and YBa$_{2}$Cu$_{4}$O$_{8}$ blocks of
normal and superconducting Y$_{2}$Ba$_{4}$Cu$_{7}$O$_{15}$. The data
confirm our previous results that adjacent CuO$_{2}$ planes have
different doping levels and that these planes are strongly coupled. --
The static spin susceptibility at the anti-ferromagnetic wave vector
exhibits a Curie--Weiss like temperature dependence in the normal
state. --
The Y$_{2}$Ba$_{4}$Cu$_{7}$O$_{15}$ data are incompatible with a phase
diagram based on a {\em single} CuO$_2$ plane theory and suggest that
the appearance of a spin gap implies interplane coupling. Additional
data for YBa$_{2}$Cu$_{4}$O$_{8}$ and YBa$_{2}$Cu$_{3}$O$_{6.982}$ are
in accord with the single plane theory. --
The temperature dependence of $T_{2G,ind}$ below T$_c$  excludes
{\em isotropic} $s$ wave superconductivity in all three compounds.
\end{abstract}
\twocolumn
\narrowtext
\section{INTRODUCTION}
\label{sec:intro}

Recently, we reported in detail nuclear magnetic resonance (NMR) and
nuclear quadrupole resonance (NQR) measurements in the
Y$_{2}$Ba$_{4}$Cu$_{7}$O$_{15}$ compound, which can be considered as
a natural multilattice consisting of alternating
YBa$_{2}$Cu$_{4}$O$_{8}$ (\mbox{1--2--4} for short) and
YBa$_{2}$Cu$_{3}$O$_{7}$ (\mbox{1--2--3}) blocks. \cite{Stern94}
The CuO${_2}$ planes of adjacent \mbox{1--2--3} and \mbox{1--2--4} blocks
form double planes whose individual planes are inequivalent and
distinguishable by NMR/NQR. By comparing  NQR frequency,
spin--lattice relaxation time, $T_1$, and magnetic shift, $K$, of
the distinct plane Cu sites we found three major results:
(i) the individual planes of a double plane have different doping levels;
(ii) the individual planes of a double plane are strongly coupled with an
 estimated coupling strength of at least 30~meV; (iii) the electron spin
fluctuation spectrum in both individual planes exhibits a spin pseudo--gap
with a common value.

To further study the interplane coupling and its consequences on
inplane spin dynamics we have performed measurements of an additional
NMR/NQR quantity, the Gaussian contribution $T_{2G}$ to
the plane Cu nuclear spin--spin relaxation time $T_{2}$.
Complementary to $K$ and $T_1$, $T_{2G}$ delivers information on
the real part of the static electronic spin susceptibility,
$\chi^{\prime}(q)$, at non--zero wave vector, $q$. \cite{Penn91}
We will present $T_{2G}$ data for Y$_{2}$Ba$_{4}$Cu$_{7}$O$_{15}$
over the temperature range from 15 to 400~K, and as well results of
comparative measurements for YBa$_{2}$Cu$_{4}$O$_{8}$ and
YBa$_{2}$Cu$_{3}$O$_{6.982}$.

The paper is organized as follows. The next section
contains some necessary theoretical background. Experimental
procedures, including the characterization of the sample, are given
in Sec.~\ref{sec:exp}. In Sec.~\ref{sec:data} we present our data,
followed by a discussion of results in Sec.~\ref{sec:disc} and a
summary in Sec.~\ref{sec:summ}.

\section{THEORY}
\label{sec:th}

Pennington {\it et al.} \cite{Pen189} were the first to realize that
the spin--spin relaxation rate of plane Cu in
YBa$_{2}$Cu$_{3}$O$_{7}$ is much larger as expected from conventional
nuclear dipolar coupling. They showed that the predominant part of
the rate is due to an enhanced Cu nuclear--nuclear spin coupling
induced through an indirect coupling via electron spins.

If the quantization axis of the Cu nuclear spin is parallel to the
crystallographic $c$ axis, as in the case of plane copper in pure NQR,
then $T^{-1}_{2G,ind}$ can be expressed \cite{Penn91,ThTa94}
in terms of $\chi^{\prime}(q)$ as
\begin{eqnarray}
\label{Theo1}
\left[\frac{1}{T_{2G,ind}}\right]^{2}
&& = \frac{P(\gamma_n \hbar)^4}{m \hbar^2}
\Biggl[\frac{1}{N}\sum_q |A(q)_{cc}|^4 \chi^{\prime}(q)^2  \nonumber \\
&& - \biggl[\frac{1}{N}\sum_q |A(q)_{cc}|^2
\chi^{\prime}(q)\biggr]^{2}\Biggr] \ .
\end{eqnarray}
\noindent
Here, $P$ and $\gamma_n$ are the abundance and the gyromagnetic ratio
of the Cu isotope being studied, $m$ is
a constant that depends on the resonance method used ($m = 8$
for NMR and $4$ for NQR), $N$ is the number of Cu atoms per unit area,
$c$ denotes the direction of quantization, {\it i.e.} the
direction of the main component of the electric field gradient tensor
in case of NQR. ${\bf A}(q)$ is the Fourier transform of the
hyperfine coupling tensor ${\bf A}({\bf r}_i)$ consisting of the
on--site $A_{cc}$ (${\bf r}_j$~=~0) and isotropic transferred $B$
(${\bf r}_j$~$\not=$~0) terms. For the Cu nuclei under consideration
and adopting the Mila--Rice Hamiltonian, \cite{Mila89} $A(q)_{cc}$ is
given by
\begin{equation}
\label{Theo2}
A(q)_{cc} = A_{cc} + 2B\left[\cos(q_{x}a)+\cos(q_{y}a)\right] \ .
\end{equation}
\noindent
Since in all Y-Ba-Cu-O compounds the spin part of the plane Cu magnetic
shift in $c$ direction is zero, $A_{cc}~= -~4B$.
Consequently, ${\bf A}(q)$ peaks at the corners of the first
Brillouin zone at $Q_{AF}~= (\pm\frac{\pi}{a}, \pm\frac{\pi}{a})$, and
$T^{-1}_{2G,ind}$ therefore involves predominantly
$q$ summation of $\chi^{\prime}(q)$ close to $Q_{AF}$.

Using a phenomenological expression \cite{MMPi90} for
$\chi^{\prime}(q)$, Thelen and Pines showed \cite{ThTa94} that, in the
long correlation length limit $\xi \gg a$,
\begin{equation}
\label{Theo3}
\frac{1}{T_{2G,ind}} \propto \frac{\chi^{\prime}(Q_{AF})}{\xi} \propto
\xi \sqrt{\beta} \ ,
\end{equation}
\noindent
where $a$ is the lattice constant and $\beta$ a parameter measuring the
relative strength of the antiferromagnetic (AF) fluctuations with respect
to the zone--center fluctuations.

Finally, we recall the ``spin--lattice relaxation rate per
temperature unit", $(T_1T)^{-1}$, which is given by the weighted
$q$ average of the
$\lim_{\omega\to 0} \frac{\chi^{\prime \prime}(q,\omega)}{\omega}$.
For $\xi \gg a$, one obtains \cite{ThTa94,MMPi90}
\begin{equation}
\label{Theo4}
\frac{1}{T_{1}T} \propto
\frac{\chi^{\prime}(Q_{AF})}{\Gamma_{AF}} \sqrt{\beta} \ ,
\end{equation}
\noindent
where $\Gamma_{AF}$ is a characteristic AF spin fluctuation energy
scale.

\section{EXPERIMENT}
\label{sec:exp}

The Y$_{2}$Ba$_{4}$Cu$_{7}$O$_{15}$ sample used in this work
was synthesized with Ba metal (not carbonate) by ``method II"
of Ref.~ \onlinecite{Karp94}. The exact oxygen content
determined by volumetric method \cite{Cond89} is $14.970(2)$, the
lattice parameters are $a = 3.8314$~\AA, $b = 3.881$~\AA \ and
$c = 50.679$~\AA, and the onset of the superconducting transition
occurs at T$_{c} = 93.1$~K.

Since the plane Cu NQR lines of the \mbox{1--2--3} and \mbox{1--2--4}
blocks are quite narrow and well separated, \cite{Karp94} we
measured $T_{2G}$ by the NQR spin echo method in zero magnetic field
using standard pulsed spectrometers. The signals were obtained by
a phase alternating add--subtract spin--echo technique
similar to that one used in Ref.\ \onlinecite{PennTh}.
To have an optimal filling factor and thus optimal signal--to--noise
ratio, we used unoriented powder, for which $T_{2G}$ is about
7$\%$ larger than for uniaxially aligned powder. \cite{Imai93}
To be able to make unbiased comparison of
Y$_{2}$Ba$_{4}$Cu$_{7}$O$_{15}$ $T_{2G}$ results with
those from its parent compounds, we in addition measured $T_{2G}$ on
unoriented YBa$_{2}$Cu$_{4}$O$_{8}$ and YBa$_{2}$Cu$_{3}$O$_{6.982}$
powders under the same experimental conditions.

In order to measure $T_{2G}$ properly it is necessary to uniformly
flip all nuclear spins involved in the experiment. This demands a
flipping pulse ($\pi$ pulse) that is  short compared to the
inverse of the linewidth. In going from normal to superconducting
state the pulse length has to be readjusted due to a changed
penetration depth and an inevitable increase of the inhomogeneity of
the flipping high--frequency field in the superconducting powder
grains. Failing to do so results in an  abrupt prolongation
of $T_{2G}$ just below T$_{c}$. In general, an incomplete flipping
of the spins due to whatsoever experimental shortcomings, results
in a prolonged $T_{2G}$, therefore all the presented
$T^{-1}_{2G}$ data are somewhat smaller than the proper value
obtained under ``ideal experimental conditions". We estimate that
for the superconducting state our $T^{-1}_{2G}$ values could be up to
20$\%$ too low because of the use of unoriented powder and the
inhomogeneous high--frequency field.

The full linewidth at half height (FWHH) of the  plane $^{63}$Cu NQR
at 100~K is $350$~kHz in the \mbox{1--2--3} and $220$~kHz in the
\mbox{1--2--4} block of Y$_{2}$Ba$_{4}$Cu$_{7}$O$_{15}$
(Ref.~\onlinecite{Karp94}), and $180$~kHz in YBa$_{2}$Cu$_{4}$O$_{8}$
(Ref.~\onlinecite{Zimm89}) and $200$~kHz in
YBa$_{2}$Cu$_{3}$O$_{6.982}$ (Ref.~\onlinecite{Schi89}).
At temperatures above T$_{c}$, the length of the applied $\pi$ pulse
was 1.4 $\mu$s, below T$_{c}$ this length increased to 2.4~$\mu$s.

\section{RESULTS AND ANALYSIS}
\label{sec:data}

The spin--echo amplitude $E$, recorded as a function of
time $\tau$ between the first and the second (flipping) pulse
could be fitted to the expression

\begin{equation}
\label{T2G}
E(2 \tau) = E_{0} \;\exp \left[-\frac{2 \tau}{T_{2 R}} - \frac{1}{2}
\left(\frac{2 \tau}{T_{2 G}}\right)^{2}\right] \ ,
\end{equation}
\noindent
where the Lorentzian--Redfield term  $T^{-1}_{2R}$ stands for the
decay rate due to the spin--lattice relaxation process.

$T^{-1}_{2R}$ was determined from the expression
$T^{-1}_{2R} = (2 + r)/3 T^{-1}_{1}$ (Ref.\ \onlinecite{Penn91}),
using NQR $T_{1}$ from Ref.\ \onlinecite{Stern94,Zimm89,Pen289}.
For the anisotropy of the relaxation rate, $r$, we took the values
$3.7$ (Ref.\ \onlinecite{Barr91}) for YBa$_{2}$Cu$_{3}$O$_{6.982}$
and the \mbox{1--2--3} block and $3.3$ (Ref.\ \onlinecite{Bank92})
for the \mbox{1--2--4} block and YBa$_{2}$Cu$_{4}$O$_{8}$, assuming
that $r$ is the same in the parent compound and in the corresponding
block of Y$_{2}$Ba$_{4}$Cu$_{7}$O$_{15}$. With $T_{2G}$ and
$T_{2R}$ as free fitting parameters we found that $T_{2R}$
thus obtained agrees with that calculated from $T_{1}$. The
disadvantage of the latter fitting procedure is a larger
scatter of the results.

Fig.~\ref{fig:rawdata} presents our results for $T^{-1}_{2G}$
of Y$_{2}$Ba$_{4}$Cu$_{7}$O$_{15}$, YBa$_{2}$Cu$_{3}$O$_{6.982}$
and YBa$_{2}$Cu$_{4}$O$_{8}$ corrected now for ``oriented powder"
by multiplying the raw values by $1.07$. For comparison, recent
data for YBa$_{2}$Cu$_{3}$O$_{6.9}$\cite{Ima293} (multiplied
by $\sqrt{2}$ for comparison with NQR values),
YBa$_{2}$Cu$_{3}$O$_{6.98}$\cite{Itoh94}
and YBa$_{2}$Cu$_{4}$O$_{8}$\cite{Itoh92} are given. While there
is rather good agreement for the YBa$_{2}$Cu$_{4}$O$_{8}$
structures, there is a discrepancy for the YBa$_{2}$Cu$_{3}$O$_{x}$
samples either for experimental reasons or because of a
dependence of $T_{2G}$ on doping (see below).

The obtained $T^{-1}_{2G}$ encompass the temperature independent
contribution, $T^{-1}_{2G,dip}$, arising from the direct nuclear
dipole--dipole interaction and the temperature dependent
contribution, $T^{-1}_{2G,ind}$, caused
by the indirect nuclear spin--spin coupling mediated through the
AF correlated electron spins. In a Gaussian approximation,
neglecting the interference terms, both
contributions add as \cite{Itoh92}
\begin{equation}
\label{T2Graw}
T^{-2}_{2G} \approx T^{-2}_{2G,dip} + T^{-2}_{2G,ind} \ .
\end{equation}
\noindent
Anticipating a common temperature dependence of
$T^{-1}_{2G,ind}$ in both planes, we plotted $(T^{-2}_{2G})_{124}$ vs.
$(T^{-2}_{2G})_{123}$ using the temperature as an implicit parameter
(Fig.~\ref{fig:rawdata}, insert). Within the experimental scatter a
linear relationship
\begin{equation}
\label{T2Gdip}
\left(T^{-2}_{2G}\right)_{124} = A \left(T^{-2}_{2G}\right)_{123} + B
\end{equation}
\noindent
with $A~= 2.01(6)$ and $B~= - 17(8)~ms^{-2}$ is indeed
observed.
If $T^{-1}_{2G,dip}$ is equal in both planes, as one may expect on
grounds of structural similarity, the relationship
(\ref{T2Gdip}) delivers $T^{-1}_{2G,dip}~= 4.1(1.0)~ms^{-1}$,
which is close to the $4.5~ms^{-1}$ we estimate for the
Y$_{2}$Ba$_{4}$Cu$_{7}$O$_{15}$ crystal structure by using a modified
Abragam--Kambe \cite{Itoh92,AbKa53} expression for the NQR dipolar
second moment. The Y$_{2}$Ba$_{4}$Cu$_{7}$O$_{15}$ $T^{-1}_{2G,dip}$
value is somewhat smaller than the estimated $5.8$ and $5.9~ms^{-1}$
obtained for YBa$_{2}$Cu$_{3}$O$_{6.98}$
and YBa$_{2}$Cu$_{4}$O$_{8}$, respectively. \cite{Itoh92} This difference
arises from the fact that in Y$_{2}$Ba$_{4}$Cu$_{7}$O$_{15}$ the
Cu NQR frequencies in the individual planes of the double plane
differ, therefore the ``neighbor" plane having only nonresonant Cu nuclear
spins contributes much less to $T^{-1}_{2G,dip}$ as in case of
YBa$_{2}$Cu$_{3}$O$_{7}$ and YBa$_{2}$Cu$_{4}$O$_{8}$, where all planes
are equivalent.

Finally, Eq.(\ref{T2Graw}) yields  $T_{2G,ind}$, the results are
plotted in Fig.~\ref{fig:t2ind}. Notice that
$(1/T_{2G,ind})_{124}~> (1/T_{2G,ind})_{123}$, meaning that
the planes of the \mbox{1--2--4} block are less doped than those of
the \mbox{1--2--3} block, in agreement with our earlier conclusions.
Further, $T_{2G,ind}$ in both blocks of
Y$_{2}$Ba$_{4}$Cu$_{7}$O$_{15}$ follows the same temperature
dependence above and below T$_c$, as seen from the constant
ratio $r_{T_2}~= T^{123}_{2G,ind}/T^{124}_{2G,ind}$
(Fig.~\ref{fig:t2ind}, insert).

Around 50~K, we observe in Y$_{2}$Ba$_{4}$Cu$_{7}$O$_{15}$ at both
plane Cu sites an unexpected growth of $T^{-1}_{2R}$, peaking at 52~K.
Similar peaks were detected at 87~K in YBa$_{2}$Cu$_{3}$O$_{6.982}$
and YBa$_{2}$Cu$_{3}$O$_{7}$. \cite{Itoh90} Since no such peaks
appear in YBa$_{2}$Cu$_{4}$O$_{8}$ which stands out in the
Y--Ba--Cu--O family as a stoichiometric, thermally very stable compound,
we suspect that the diffusion of loosely bound oxygen in the single
chains, known as the weak structural elements present in
YBa$_{2}$Cu$_{3}$O$_{7}$ and Y$_{2}$Ba$_{4}$Cu$_{7}$O$_{15}$ but
not in YBa$_{2}$Cu$_{4}$O$_{8}$, could be the source of this
$T_{2R}$ anomaly. A more detailed study to clarify this aspect is
necessary.  However, we believe that this anomaly is extrinsic
rather than a genuine effect of the inplane electron spin dynamics.

\section{DISCUSSION}
\label{sec:disc}

The experimental results on $T_{2G}$ allow us to discuss five
features which characterize Y$_{2}$Ba$_{4}$Cu$_{7}$O$_{15}$ at the
microscopic level: the interplane coupling, the temperature dependence of
the electronic susceptibility at the AF wave vector, the spin gap and its
relation to the interplane coupling and, finally, the symmetry of the
pair wave function of the superconducting state.

\subsection{Further evidence for interplane coupling}
\label{sec:coupling}

The presence of interplane coupling, at least for the normal conducting
phase of Y$_{2}$Ba$_{4}$Cu$_{7}$O$_{15}$,  were deduced \cite{Stern94}
from the fact that the spin--lattice relaxation rate  for the two
inequivalent plane Cu sites exhibited the {\em same} temperature
dependence; the same was found for the Knight shift at these sites.
The temperature dependence itself shows a behavior typical for an
underdoped high--T$_{c}$ compound that forms a spin gap (see below). Since
both the relaxation rate and the Knight shift are related to the dynamic
susceptibility, the common temperature dependence reveals the same
dynamics in these planes which must arise from a coupling between the
planes.

As shown by the insert of Fig.~\ref{fig:t2ind}, also the $T_{2G}^{-1}$
data of the two inequivalent Cu sites reveal a common temperature
dependence and thus provide a third piece of evidence for the interplane
coupling.

\subsection{Temperature dependence of $\chi(Q_{AF})$ above T$_c$}
\label{sec:hii}

In principle, using experimental $T_{2G}$ and $T_1$ data, the electronic
susceptibility at the AF wave vector, in the long correlation length
limit, can be derived from Eqs.~(\ref{Theo3}) and (\ref{Theo4}),
respectively. This requires, however, some knowledge about the temperature
dependence of the parameters $\xi$ and $\beta$ which is not known
{\em a priori}.
Nevertheless, similar to the treatment of YBa$_{2}$Cu$_{3}$O$_{6.9}$ in
Ref.~\onlinecite{Ima293}, we can discuss two limiting cases.
(i) If the correlation length, $\xi$, is independent of temperature,
$\chi'(T)$ is determined by the temperature dependence of $T_{2G,ind}$
according to Eq. (\ref{Theo3}), hence $\chi'(Q_{AF}) \propto 1/T_{2G,ind}$.
(ii) If on the other hand, $\xi$ depends strongly on temperature and $\beta$
does not, we have $1/T_{2G,ind} \propto \xi$ and
$\chi'(Q_{AF}) \propto \xi^2$, hence $\chi'(Q_{AF}) \propto T^{-2}_{2G,ind}$.

In Fig.~\ref{fig:curie} we have plotted the temperature dependence of both
$T_{2G,ind}$ and $T_{2G,ind}^2$ corresponding to cases (i) and (ii),
respectively, together with our data for YBa$_{2}$Cu$_{3}$O$_{6.982}$ and
YBa$_{2}$Cu$_{4}$O$_{8}$. For case (i), all four data sets can be fitted by a
Curie-Weiss law (dotted curves) which implies that $1/\chi'(Q_{AF}) \propto
1/(T + \Theta)$ where $\Theta$ is the Weiss temperature.
For case (ii), the inverse susceptibility of the \mbox{1--2--3} and the
\mbox{1--2--4} blocks of Y$_{2}$Ba$_{4}$Cu$_{7}$O$_{15}$ follows a Curie law,
$i.e.$ $1/\chi'(Q_{AF}) \propto 1/T$, while the YBa$_{2}$Cu$_{3}$O$_{7}$ and
YBa$_{2}$Cu$_{4}$O$_{8}$ data show a different behavior.

We believe that case (i), $i.e.$ a temperature independent or at least
only weakly dependent correlation length, is the correct interpretation of
the data since it applies to all four data sets with a fit quality that is
better than in case (ii).  Furthermore, this result is in accord with
theoretical models \cite{FukMor} and supports results of inelastic neutron
scattering in YBa$_{2}$Cu$_{3}$O$_{x}$. \cite{Ross92}

Accepting this conclusion, we note that the negative value of the Weiss
temperatures imply that the compounds do not order anti-ferromagnetically
at low temperatures which is a well-known fact. For the two blocks of
Y$_{2}$Ba$_{4}$Cu$_{7}$O$_{15}$, the Weiss temperatures are the same
(about --200~K) while $\Theta$ is --100~K and --300~K in the
YBa$_{2}$Cu$_{4}$O$_{8}$ and YBa$_{2}$Cu$_{3}$O$_{7}$ structures,
respectively. Thus, the absolute value of the Weiss temperature in the
Y-Ba-Cu-O compounds increases with rising doping level and distinguishes
Y$_{2}$Ba$_{4}$Cu$_{7}$O$_{15}$ as a (nearly) optimized compound.  A similar
behavior, according to $T_1$ measurements, \cite{Kita92} had been found
for La-Sr-Cu-O compounds where the Sr content determines the doping level.

\subsection{Spin gap behavior and interplane coupling}
\label{sec:spingap}

Next, we will discuss the energy scale parameter of the AF fluctuations,
$\Gamma_{AF}$, and the evidence for the presence of a spin gap. NMR
investigations of underdoped YBa$_{2}$Cu$_{3}$O$_{x}$ compounds,
\cite{BrMa94} of YBa$_{2}$Cu$_{4}$O$_{8}$ (Ref.~\onlinecite{Bank94})
and of Y$_{2}$Ba$_{4}$Cu$_{7}$O$_{15}$ (Ref.~\onlinecite{Stern94})
have shown that the planar Cu spin-lattice relaxation and Knight shift
data can be interpreted in terms of a spin gap, in accord with
neutron scattering data. \cite{Ross92} The occurrence of a spin gap
means that spectral weight in the electron spin fluctuations is
transferred from lower to higher energy. The presence of the spin gap
manifests itself in a maximum of $1/T_1T$ at a temperature T$^\ast$
well above T$_c$, with T$^\ast$ = 130 and 150~K for
Y$_{2}$Ba$_{4}$Cu$_{7}$O$_{15}$ (Ref. 1) and YBa$_{2}$Cu$_{4}$O$_{8}$
(Ref. 25), respectively. It should be stressed again that the $T^{-1}_{2G}$
data do not show such a peak.

Combining the Eq.~(\ref{Theo3}) and (\ref{Theo4}) yields the quantity
$T_1T/T_{2G,ind}^2$ which is proportional to $\Gamma_{AF}$. The plot of
Fig.~\ref{fig:t1t/t22} reveals that for both blocks of
Y$_{2}$Ba$_{4}$Cu$_{7}$O$_{15}$ and for YBa$_{2}$Cu$_{4}$O$_{8}$,
$\Gamma_{AF}$ increases with falling temperature whereas for
YBa$_{2}$Cu$_{3}$O$_{6.982}$ it remains constant. The latter result is in
agreement with earlier measurements by the Slichter group;
\cite{Ima293} our YBa$_{2}$Cu$_{4}$O$_{8}$ data are
similar to those of Itoh {\it et al.} \cite{Itoh92}

In the framework of the random phase approximation (RPA) formalism
\cite{MMPi90,Itoh94} one does not expect such an increase of
$\Gamma_{AF}$, instead the spin fluctuations are supposed to be slowed
down by growing AF correlation on decreasing temperature. However, if
the spin excitation spectrum is changed by the opening of a spin gap,
the increase of $\Gamma_{AF}$ is conceivable.

The origin of the spin gap is still under debate. For instance, one may
ask whether the gap is an intrinsic property of the single CuO${_2}$
plane or a consequence of interplane effects. One approach to answer this
question \cite{SoPi93,BaPi94} considers the spin gap an intrinsic
property of a single underdoped CuO${_2}$ plane that shows a quasi
two-dimensional quantum Heisenberg antiferromagnet behavior. The spin gap
itself is related to the suppression of spectral weight for frequencies
smaller than $v_s/\xi$ ($v_s$ is the spin wave velocity) in the quantum
disordered regime. Another approach \cite{MiMo93,Ioff94,ULee94} assumes
that the gap originates from an effective interplane coupling between
adjacent CuO${_2}$ planes.

Pursuing the first path, Sokol and Pines (SP) \cite{SoPi93} propose an
unified magnetic phase diagram of the cuprate superconductors, which,
dependent on doping level and temperature, displays various regimes that,
among others, are characterized by certain temperature independent ratios
of plane Cu $T_1T$ and $T_{2G,ind}$ values. In the {\it quantum critical}
(QC) regime (applicable to spin gap compounds), the ratio
$T_1T/T_{2G,ind}$ is constant while in the {\it overdamped} (OD) regime
$T_1T/T_{2G,ind}^2$ is constant.

We check these predictions with the help of Fig.~\ref{fig:t1t/t22} and
its insert. Obviously, the overdoped compound YBa$_{2}$Cu$_{3}$O$_{7}$
is in the OD regime from T$_c$ up to 300~K (as noted already by SP,
although only for the 150--300 K range). On the other hand, the
underdoped YBa$_{2}$Cu$_{4}$O$_{8}$ structure is in the QC regime for
temperatures above T$_{124}^\ast$.

While these results are in accord with the SP model, the
Y$_{2}$Ba$_{4}$Cu$_{7}$O$_{15}$ structure does not fit into this scheme.
The mere fact that the differently doped planes of the \mbox{1--2--3} and
\mbox{1--2--4} blocks have the same cross-over temperature T$_{247}^\ast$
already contradicts the proposed phase diagram. In particular, above
T$_{247}^\ast$, the two blocks do not show the
$T_1T/T_{2G,ind} =\ constant$ behavior expected for the QC regime the
material should belong to because of the spin gap. Instead, we note that
$T_1T/T^{2}_{2G,ind} =\ constant$ which is the signature of the OD
regime.

On the other hand, the second approach, by taking the interplane coupling
as the origin of the spin gap, quite naturally accounts for the
observed Y$_{2}$Ba$_{4}$Cu$_{7}$O$_{15}$ behavior. The interplane coupling
not only dictates a common temperature dependence of the dynamic
susceptibility in adjacent planes, it also leads to a {\em common} value
of the spin gap in the fluctuation spectrum. If we make the reasonable
assumption that the effective interplane coupling depends on the sum of
charge carrier effects in both planes, one understands why the
cross--over temperature of Y$_{2}$Ba$_{4}$Cu$_{7}$O$_{15}$,
T$^\ast$ = 130 K, lies somewhere between the corresponding values for
YBa$_{2}$Cu$_{3}$O$_{7}$ (T$^\ast \le$ T$_c$) and
YBa$_{2}$Cu$_{4}$O$_{8}$ (T$^\ast$ = 150 K).

\subsection{Symmetry of superconducting state}
\label{sec:sc}

We now discuss the relevance of our results with respect to the
symmetry of the pair wave function of the superconducting state. We
therefore have plotted in Fig.~\ref{fig:belowTc}
$T^{-1}_{2G,ind}($T$)/T^{-1}_{2G,ind}($T$_{c}$) vs T/T$_{c}$ for all plane
Cu sites in Y$_{2}$Ba$_{4}$Cu$_{7}$O$_{15}$, YBa$_{2}$Cu$_{4}$O$_{8}$
and YBa$_{2}$Cu$_{3}$O$_{6.982}$.
Within error limits the various data points gather on an ``universal"
curve which decreases monotonously towards $0.80$ in the T$\to 0$ limit.
According to  the discussion of Sec.~\ref{sec:exp}, the ``true" limiting
value of $T^{-1}_{2G,ind}($T$)/T^{-1}_{2G,ind}($T$_{c}$) could be
about 20$\%$ {\em higher}.

The dashed and dotted lines in Fig.~\ref{fig:belowTc} are theoretical
curves calculated by Bulut and Scalapino \cite{Bulu91} for
YBa$_{2}$Cu$_{3}$O$_{7}$ in case of $s$ and $d$ wave symmetry. Although
the corresponding calculations for Y$_{2}$Ba$_{4}$Cu$_{7}$O$_{15}$ and
YBa$_{2}$Cu$_{4}$O$_{8}$ have not yet been performed it is reasonable to
assume that these ``normalized" curves will be not much different from the
YBa$_{2}$Cu$_{3}$O$_{7}$ curves since the experimental data for all
compounds follow an universal curve.

The experimental points are much closer to the $d$ than the $s$ wave
curve, in particular if a possible 20$\%$ correction is taken into
account. Hence, the $T_{2G,ind}$ data seem to favor $d$ wave symmetry.
For YBa$_{2}$Cu$_{4}$O$_{8}$, this result is in accord with our previous
conclusion drawn from Cu spin--lattice relaxation and Knight shift
measurements \cite{Bank94}. However, as shown by Sudb{\o} {\it et
al.}, \cite{Sudb94} models of $d$ wave and strongly {\it anisotropic} $s$
wave superconductivity deliver very similar results. Thus, we conclude
that the $T_{2G}$ data  exclude {\em isotropic} $s$ wave symmetry.

\section{SUMMARY}
\label{sec:summ}
We have presented plane Cu nuclear spin--spin relaxation rates,
$T^{-1}_{2}$, of Y$_{2}$Ba$_{4}$Cu$_{7}$O$_{15}$,
YBa$_{2}$Cu$_{4}$O$_{8}$ and YBa$_{2}$Cu$_{3}$O$_{6.982}$, which
deliver information on the static electron spin susceptibility,
$\chi^{\prime}(q)$, at non--zero wave vector, $q$. Our main conclusions
were drawn from a discussion of the indirect component, $T_{2G,ind}$, of
the Gaussian contribution to $T_2$.

For Y$_{2}$Ba$_{4}$Cu$_{7}$O$_{15}$, the magnitude of the individual
$T_{2G,ind}$ values confirms our previous conclusion \cite{Stern94} that
the planes of the \mbox{1--2--4} block are less doped than the
\mbox{1--2--3} block planes. The temperature dependence of $T_{2G,ind}$
provides further evidence for a strong interplane coupling between
adjacent CuO$_2$ planes belonging to different blocks.

For all three compounds studied, the data suggest a nearly  temperature
independent correlation length of the anti-ferromagnetic (AF) fluctuations
and a Curie--Weiss law for the susceptibility at the AF wave vector with
negative Weiss temperatures whose absolute values increase with doping
level.

The data support our previous conclusion about the existence of the spin
pseudo--gap in Y$_{2}$Ba$_{4}$Cu$_{7}$O$_{15}$ and YBa$_{2}$Cu$_{4}$O$_{8}$.
The observation of the same gap value for both
blocks in Y$_{2}$Ba$_{4}$Cu$_{7}$O$_{15}$ points to the importance of the
interplane coupling in the gap formation. We show that the
Y$_{2}$Ba$_{4}$Cu$_{7}$O$_{15}$ results are incompatible with the
Sokol---Pines phase diagram \cite{SoPi93} based on single CuO$_2$ plane
theory while the YBa$_{2}$Cu$_{4}$O$_{8}$ and
YBa$_{2}$Cu$_{3}$O$_{6.982}$ data are in accord with that theory.

The temperature dependence of $T_{2G,ind}$ below T$_c$ excludes
{\em isotropic} $s$ wave superconductivity in
Y$_{2}$Ba$_{4}$Cu$_{7}$O$_{15}$, YBa$_{2}$Cu$_{4}$O$_{8}$ and
YBa$_{2}$Cu$_{3}$O$_{6.982}$.
Together with our spin--lattice relaxation and Knight shift measurements
in YBa$_{2}$Cu$_{4}$O$_{8}$ (Ref.~\onlinecite{Bank94}) and corresponding
investigations of YBa$_{2}$Cu$_{3}$O$_{7}$ (Ref.~\onlinecite{Slic93}), one
may conclude that all
Y-Ba-Cu-O compounds are not {\em isotropic} $s$ wave superconductors.
At present it seems impossible to distinguish between possible $d$ and
more exotic $s$ wave superconductivity in these materials.

\acknowledgments
We thank the group of Prof. E. Kaldis (ETH--Zurich) for preparing the
Y$_{2}$Ba$_{4}$Cu$_{7}$O$_{15}$ material. Financial support by the Swiss
National Science Foundation is gratefully acknowledged.

\begin{figure}
\caption{Temperature dependence of NQR $T^{-1}_{2G}$ at the plane copper
sites in the \mbox{1--2--3} ($\circ$) and \mbox{1--2--4} block
($\bullet$) of Y$_{2}$Ba$_{4}$Cu$_{7}$O$_{15}$, in
YBa$_{2}$Cu$_{3}$O$_{6.982}$ (\protect\raisebox{-.3ex}{$\Box$}), and
YBa$_{2}$Cu$_{4}$O$_{8}$ (\protect\rule{1.6mm}{1.6mm}).
For comparison:  data for
YBa$_{2}$Cu$_{3}$O$_{6.9}$ ($\triangle$, joined with dotted line,
Ref.\ \protect\onlinecite{Ima293}), YBa$_{2}$Cu$_{3}$O$_{6.98}$
(dash--dotted line, Ref.\ \protect\onlinecite{Itoh94}) and
YBa$_{2}$Cu$_{4}$O$_{8}$ (dashed line, Ref.\ \protect\onlinecite{Itoh92}).
Insert: $T^{-2}_{2G}$ of \mbox{1--2--4} block vs that of
\mbox{1--2--3} block with temperature as an implicite parameter.}
\label{fig:rawdata}
\end{figure}

\begin{figure}
\caption{Temperature dependences of NQR $T^{-1}_{2G,ind}$
at the plane copper sites. The symbols are the same as in
Fig.~\protect\ref{fig:rawdata}.
Insert: ratio of $T_{2G,ind}$ of individual planes in
Y$_{2}$Ba$_{4}$Cu$_{7}$O$_{15}$.}
\label{fig:t2ind}
\end{figure}

\begin{figure}
\caption{Temperature dependences of (a) $T_{2G,ind}$
and (b) $T^{2}_{2G,ind}$.The symbols are the same as in
Fig.~\protect\ref{fig:rawdata}. Dotted lines are fits to the (a)
Curie--Weiss or (b) Curie law.}
\label{fig:curie}
\end{figure}

\begin{figure}
\caption{Temperature dependences of
$T_1T/T^{2}_{2G,ind}$ ($\propto \Gamma_{AF}$) and $T_1T/T_{2G,ind}$
(insert). The symbols are the same as in Fig.~\protect\ref{fig:rawdata},
the lines are guides to the eye, the arrows mark T$_{c}$ and $T^{\ast}$
of Y$_{2}$Ba$_{4}$Cu$_{7}$O$_{15}$.}
\label{fig:t1t/t22}
\end{figure}

\begin{figure}
\caption{Temperature dependences of $T^{-1}_{2G,ind}$
at the plane copper sites in the superconducting state. The symbols are
the same as in Fig.~\protect\ref{fig:rawdata}, the solid line is a
guide to the eye. The dashed and dotted lines are calculated values
(Ref.~\protect\onlinecite{Bulu91})
for a $s$ and a $d$ wave superconductor, respectively.}
\label{fig:belowTc}
\end{figure}

\begin{references}
\bibitem[*]{Auth1}      Also at the Institute of Chemical Physics and
                        Biophysics, EE--0001 Tallinn, Estonia.
\bibitem{Stern94}       R.\ Stern,
                        I.\ Mangelschots, M.\ Mali, J.\ Roos,
                        D.\ Brinkmann, J.--Y.\ Genoud, T.\ Graf, and J.\
Muller,
                        Phys.\ Rev.\ B.\ {\bf 50}, 426 (1994).
\bibitem{Penn91}        C.H. Pennington and C.P. Slichter,
                        Phys.\ Rev.\ Lett.\ {\bf 66}, 381 (1991).
\bibitem{Pen189}        C.\ H.\ Pennington, D.\ J.\ Durand, C.\ P.\ Slichter,
                        J.\ P.\ Rice, E.\ D.\ Bukowski, and D.\ M.\ Ginsberg,
                        Phys.\ Rev.\ B.\ {\bf 39}, 274 (1989).
\bibitem{ThTa94}        (a) D. Thelen and D. Pines, Phys.\ Rev.\ B.\ {\bf 49},
                        3528 (1994); (b) M. Takigawa, {\it ibid.}, 4158 (1994).
\bibitem{Mila89}        F.\ Mila and T.\ M.\ Rice,
                        Physica C {\bf 157}, 561 (1989).
\bibitem{MMPi90}        A.\ J.\ Millis, H.\ Monien, and D.\ Pines,
                        Phys.\ Rev.\ B.\ {\bf 42}, 167 (1990).
\bibitem{Karp94}        J.\ Karpinski, K.\ Conder, H.\ Schwer, Ch.\ Kr\"{u}ger,
                        E.\ Kaldis, M.\ Maciejewski, C.\ Rossel, M.\ Mali, and
                        D.\ Brinkmann,
                        Physica C {\bf 227}, 68 (1994).
\bibitem{Cond89}        K.\ Conder, S.\ Rusiecki, and E.\ Kaldis,
                        Mater. Res. Bull. {\bf 24}, 581 (1989).
\bibitem{PennTh}        C.H. Pennington, PhD Theses,
                        University of Illinois (1989).
\bibitem{Imai93}        T. Imai, C.\ P Slichter, K.\ Yoshimura, M.\ Katoh, and
                        K.\ Kosuge,
                        Phys.\ Rev.\ Lett.\ {\bf 71}, 1254 (1993).
\bibitem{Zimm89}        H.\ Zimmermann, M.\ Mali, D.\ Brinkmann,
                        J.\ Karpinski, E.\ Kaldis, and S.\ Rusiecki,
                        Physica C {\bf 159}, 681 (1989).
\bibitem{Schi89}        H.\ Schiefer, M.\ Mali, J.\ Roos, H.\ Zimmermann,
                        D.\ Brinkmann, S.\ Rusiecki, and E.\ Kaldis,
                        Physica C {\bf 162--164}, 171 (1989).
\bibitem{Pen289}        C.\ H.\ Pennington, D.\ J.\ Durand, C.\ P.\ Slichter,
                        J.\ P.\ Rice, E.\ D.\ Bukowski, and D.\ M.\ Ginsberg,
                        Phys.\ Rev.\ B.\ {\bf 39}, 2902 (1989).
\bibitem{Barr91}        S.\ E.\ Barrett, J.\ A.\ Martindale, D.\ J.\ Durand,
                        C.\ H.\ Pennington, C.\ P.\ Slichter,
                        T.\ A.\ Friedmann, J.\ P.\ Rice, and D.\ M.\ Ginsberg,
                        Phys.\ Rev.\ Lett.\ {\bf 66}, 108 (1991).
\bibitem{Bank92}        M.\ Bankay, M.\ Mali, J.\ Roos, I.\ Mangelschots, and
                        D.\ Brinkmann,
                        Phys.\ Rev.\ B.\ {\bf 46}, 11228 (1992);
                        H.\ Zimmermann, M.\ Mali, M.\ Bankay, and D.\
Brinkmann,
                        Physica C {\bf 185--189}, 1145 (1991).
\bibitem{Ima293}        T.\ Imai, C.\ P.\ Slichter, A.\ P.\ Paulikas, and
						B.\ Veal,
                        Phys.\ Rev.\ B.\ {\bf 47}, 9158 (1993);
                        Appl. Magn. Reson. {\bf 3}, 729 (1992).
\bibitem{Itoh94}        Y.\ Itoh, K.\ Yoshimura, T.\ Ohomura, H.\ Yasuoka,
                        Y.\ Ueda, and K.\ Kosuge,
                        J.\ Phys.\ Soc.\ Jpn.\ {\bf 63}, 1455 (1994).
\bibitem{Itoh92}        Y. Itoh, H.\ Yasuoka, Y.\ Fujiwara, Y.\ Ueda, T.\
Machi,
                        I.\ Tomeno, K.\ Tai, N.\ Koshizuka, and S.\ Tanaka,
                        J.\ Phys.\ Soc.\ Jpn.\ {\bf 61}, 1287 (1992).
\bibitem{AbKa53}        A.\ Abragam and K.\ Kambe,
                        Phys.\ Rev.\ {\bf 91}, 894 (1953).
\bibitem{Itoh90}        Y. Itoh, H.\ Yasuoka, and Y.\ Ueda,
                        J.\ Phys.\ Soc.\ Jpn.\ {\bf 59}, 3463 (1990).
\bibitem{FukMor}        T.\ Moriya, Y.\ Takahashi, and K.\ Ueda,
                        J.\ Phys.\ Soc.\ Jpn.\ {\bf 59}, 2905 (1990);
                        T.\ Tanamoto, K.\ Kuboki, and H.\ Fukuyama,
                        {\it ibid.} {\bf 60}, 3072 (1991).
\bibitem{Ross92}        J.\ Rossat--Mignod, L.\ P.\ Regnault, P.\ Bourges,
                        P.\ Burlet, C.\ Vettier, and J.\ Y.\ Henry,
                        in {\it Selected Topics in Superconductivity}, edited
                        by L.\ C.\ Gupta and M.\ S.\ Multani (World Scientific,
                        Singapore, 1993), p.\ 265;
                        Physica B {\bf 199\&200}, 281 (1994).
\bibitem{Kita92}        Y.\ Kitaoka, K.\ Ishida, S.\ Oshugi, K.\ Fujiwara,
                        G.--q.\ Zheng, and K.\ Asayama,
                        Appl. Magn. Reson. {\bf 3}, 549 (1992).
\bibitem{BrMa94}        D. Brinkmann and M. Mali,
                        in {\it NMR - Basic Principles and Progress},
                        ed. by P. Diehl, E. Fluck, H. G\"unther, R. Kosfeld,
                        J. Seelig, vol. 31, p. 171 (Springer, Berlin, 1994).
\bibitem{Bank94}        M.\ Bankay, M.\ Mali, J.\ Roos, and D.\ Brinkmann,
                        Phys.\ Rev.\ B.\ {\bf 50}, 6416 (1994).
\bibitem{SoPi93}        A.\ Sokol and D.\ Pines,
                        Phys.\ Rev.\ Lett.\ {\bf 71}, 2813 (1993).
\bibitem{BaPi94}        V.\ Barzykin, D.\ Pines, A.\ Sokol, and D. Thelen,
                        Phys.\ Rev.\ B.\ {\bf 49}, 1544 (1994).
\bibitem{MiMo93}        A.\ J.\ Millis and  H.\ Monien,
                        Phys.\ Rev.\ Lett.\ {\bf 70}, 2810 (1993);
                        {\bf 71} 210(E) (1993); Phys.\ Rev.\ B., in press.
\bibitem{Ioff94}        L.\ B.\ Ioffe, A.\ I.\ Larkin, A.\ J.\ Millis, and
                        P.\ L.\ Altshuler,
                        JETP\ Lett.\ {\bf 59}, 65 (1994).
\bibitem{ULee94}        M.\ U.\ Ubbens and P.\ A.\ Lee,
                        Phys.\ Rev.\ B.\ {\bf 50}, 438 (1994).
\bibitem{Bulu91}        N. Bulut and D.J. Scalapino,
                        Phys.\ Rev.\ Lett.\ {\bf 67}, 2898 (1991).
\bibitem{Sudb94}        A.\ Sudb{\o}, S.\ Chakravarty, S.\ Strong, and
                        P.\ W.\ Anderson,
                        Phys.\ Rev.\ B.\ {\bf 49}, 12245 (1994).
\bibitem{Slic93}        C.\ P.\ Slichter, J.\ A.\ Martindale, S.\ E.\ Barrett,
                        K.\ E.\ O'Hara, S.\ M.\ De Soto, T.\ Imai,
                        D.\ J.\ Durand, C.\ H.\ Pennington, T.\ A.\ Friedmann,
                        W.\ C.\ Lee, and D.\ M.\ Ginsberg,
                        J.\ Phys.\ Chem.\ Solids\ {\bf 54}, 1439 (1993).
\end{references}
\end{document}